\documentclass[lettersize,journal]{IEEEtran}
\usepackage{amsmath,amsfonts}
\usepackage{algorithmic}
\usepackage{algorithm}
\usepackage{array}
\usepackage[caption=false,font=normalsize,labelfont=sf,textfont=sf]{subfig}
\usepackage{textcomp}
\usepackage{stfloats}
\usepackage{url}
\usepackage{verbatim}
\usepackage{graphicx}
\usepackage{cite}
\usepackage{makecell}
\usepackage{float}
\begin{document}

\title{FPGA Design and Implementation of Fixed-Point Fast Divider\\ Using Goldschmidt Division Algorithm and Mitchell\\ Multiplication Algorithm}

\author{Jinkun Yang~\IEEEmembership{}
\thanks{}
\thanks{Jinkun Yang is with the School of Integrated Circuit Science and Engineering (Exemplary School of Microelectronics), University of Electronic Science and Technology of China, Chengdu, 611731, China (e-mail: 2022040906025@std.uestc.edu.cn).}}

\markboth{Manuscript, August~2025}%
{Shell \MakeLowercase{\textit{et al.}}: FPGA Design and Implementation of Fixed-Point Fast Divider Using Goldschmidt Division Algorithm and Mitchell Multiplication Algorithm}

\IEEEpubid{}

\maketitle

\begin{abstract}
This paper presents a variable bit-width fixed-point fast divider using Goldschmidt division algorithm and Mitchell multiplication algorithm. Described using Verilog HDL and implemented on a Xilinx XC7Z020-2CLG400I FPGA, the proposed divider achieves over 99\% computational accuracy with a minimum latency of 99.1 ns, which is 31.7 ns faster than existing single-precision dividers. Compared with a Goldschmidt divider using a Vedic multiplier, the proposed design reduces Slice Registers by 46.68\%, Slice LUTs by 4.93\%, and Slices by 11.85\%, with less than 1\% accuracy loss and only 24.1 ns additional delay. These results demonstrate an improved balance between computational speed and resource utilization, making the divider well-suited for high-performance FPGA-based systems with strict resource constraints. 
\end{abstract}


\section{Introduction}
\IEEEPARstart{H}{ardware} dividers implemented on field-programmable gate arrays (FPGAs) are critical components in many digital integrated circuit systems. Traditional fixed-point and floating-point dividers suffer from large computational latency and inflexible input bit-width. Function iteration methods provide an effective approach to improving divider performance. Among them, Newton–Raphson method \cite{ref1} and Goldschmidt division algorithm \cite{ref2} are widely adopted. Since the multiplications in Newton–Raphson method must be performed sequentially, while Goldschmidt algorithm achieves faster convergence with inherently parallel multiplications, the latter offers significant advantages in hardware implementations.

This paper proposes a variable bit-width fixed-point fast divider using Goldschmidt division algorithm and Mitchell multiplication algorithm \cite{ref4}. By employing Mitchell algorithm in the multiplication units of the Goldschmidt iterations, the proposed divider achieves higher computational speed with modest accuracy loss, while maintaining relatively low resource utilization. Furthermore, the input and output bit-widths of the divider can be configured through parameters, enabling more flexible applications. Section II of this paper introduces the principles of Goldschmidt division algorithm and Mitchell multiplication algorithm. Section III presents the FPGA-based implementation of the proposed divider. Section IV provides the on-board operation results and performance analysis. Section V concludes the paper.

\section{The Principles of Algorithms}
\subsection{Goldschmidt division algorithm}
Goldschmidt division algorithm \cite{ref2} is an iterative method based on multiplication and addition. Given a dividend $a$ and divisor $b$, the quotient is expressed as $q=a/b$. Letting $b=1-y$, the expression can be rewritten as:
\begin{align}
q&=\frac{a}{b}=\frac{a}{1-y}=\frac{a\left(1+y\right)}{\left(1-y\right)\left(1+y\right)}=\frac{a\left(1+y\right)\left(1+y^2\right)}{\left(1-y^2\right)\left(1+y^2\right)} \notag \\
&=\frac{a\left(1+y\right)\left(1+y^2\right)\cdots\left(1+y^s\right)}{1-y^s}. \label{deqn_ex1}
\end{align}

\noindent From equation (\ref{deqn_ex1}), if the convergence condition \begin{math}b\in[1/2,1)\end{math} is satisfied, the denominator \begin{math}b_s=1-y^s\end{math} approaches 1 as the iteration count $s$ increases, while the numerator \begin{math}a_s=a(1+y)(1+y^2)\cdots(1+y^s)\end{math} gradually converges to $q$. Based on this principle, the quotient can be approximated. At each iteration, the iteration coefficient is defined as \begin{math}m_s=1+y^s=2-b_{s-1}\end{math}. The updated numerator and denominator are given by:
\begin{subequations}\label{eq:2}
\begin{align}
a_s&=m_sa_{s-1} \label{eq:2A},\\
b_s&=m_sb_{s-1} \label{eq:2B}.
\end{align}
\end{subequations}

\noindent The approximation accuracy of \begin{math}a_s\end{math} improves exponentially with $s$. The error between the approximate quotient \begin{math}a_s\end{math} and the exact quotient $q$ can be expressed by:
\begin{equation}
\label{deqn_ex3}
q-a_s\le2^{-2s}q.
\end{equation}

\noindent Equation (\ref{deqn_ex3}) indicates that after $s$ iterations, the precision of \begin{math}a_s\end{math} reaches \begin{math}2^s\end{math} bits. Each iteration of the Goldschmidt division algorithm involves only one subtraction and two parallel multiplications. Due to its fast convergence and parallelism, the algorithm enables high-speed division.

\subsection{Mitchell approximate multiplication algorithm}
Mitchell multiplication algorithm \cite{ref4} is an approximate binary multiplication method that replaces multiplication with addition. In traditional shift-based multipliers, both the number of computational steps and hardware resource usage increase with the growth of the input bit-width. In contrast, Mitchell algorithm maintains a fixed number of computational steps, thereby significantly improving multiplication speed while requiring relatively fewer hardware resources. This advantage becomes more obvious for long input bit-width, making this algorithm particularly suitable for constructing multipliers in Goldschmidt iteration units.

For a decimal number $N(N>1)$, its binary representation can be written as:
\begin{equation}
\label{deqn_ex4}
N=z_kz_{k-1}\cdots z_1z_0.z_{-1}z_{-2}\cdots z_j,
\end{equation}

\noindent where \begin{math}z_k=1\end{math}. Thus, $N$ can be expressed as:
\begin{equation}
\label{deqn_ex5}
N=\sum_{i=j}^{k}2^iz_i=2^k\left(1+\sum_{i=j}^{k-1}2^{i-k}z_i\right)=2^k(1+x),
\end{equation}

\noindent where
\begin{equation}
\label{deqn_ex6}
x=\sum_{i=j}^{k-1}2^{i-k}z_i=0.z_{k-1}z_{k-2}\cdots z_j.
\end{equation}

\noindent In equation (\ref{deqn_ex6}), $x$ corresponds to the fractional part of $N$ when $N$ is right-shifted by $k$ bits. Taking the base-2 logarithm on both sides of (\ref{deqn_ex5}) and using the approximation \begin{math}\log_2(1+x)\approx x\end{math}, we obtain:
\begin{equation}
\label{deqn_ex7}
\log_2N\approx k+x.
\end{equation}

\noindent Accordingly, the multiplication of two non-negative real numbers $N$ and $M$ can be approximated as:
\begin{equation}
\label{deqn_ex8}
\log_2(NM)=\log_2N+\log_2M\approx k_N+k_M+x_N+x_M.
\end{equation}

\noindent By applying the inverse transform of (\ref{deqn_ex5}) and (\ref{deqn_ex7}), the product $NM$ can be approximated as:
\begin{subequations}\label{eq:9}
\begin{align}
NM&\approx2^{k_N+k_M}(1+x_N+x_M),\ x_N+x_M<1 \label{eq:9A}\\
NM&\approx2^{k_N+k_M+1}(x_N+x_M),\ \ \ x_N+x_M\geq1 \label{eq:9B}
\end{align}
\end{subequations}

\noindent Equations (\ref{eq:9A}) and (\ref{eq:9B}) represent the Mitchell multiplication without correction terms, where the maximum relative error is $1/9$. Its accuracy is insufficient for Goldschmidt iteration.

From equation (\ref{deqn_ex5}), the precise product can be expressed as:
\begin{equation}
\label{deqn_ex10}
NM=2^{k_N+k_M}(1+x_N+x_M+x_Nx_M).
\end{equation}

\noindent Subtracting (\ref{eq:9A}) and (\ref{eq:9B}) from (\ref{deqn_ex10}) respectively yields the correction terms:
\begin{subequations}\label{eq:11}
\begin{align}
(\ref{deqn_ex10})-(\ref{eq:9A})&=2^{k_N+k_M}(x_Nx_M),\ \ x_N+x_M<1 \label{eq:11A}\\
(\ref{deqn_ex10})-(\ref{eq:9B})&=2^{k_N+k_M}\left(x_N^\prime x_M^\prime\right),\ x_N+x_M\geq1 \label{eq:11B}
\end{align}
\end{subequations}

\noindent where \begin{math}x_N^\prime=1-x_N\end{math} and \begin{math}x_M^\prime=1-x_M\end{math}. Adding the correction terms to (\ref{eq:9A}) and (\ref{eq:9B}), the exact product can be obtained. For hardware design convenience, let \begin{math}k_{NM}\end{math} be the integer part of \begin{math}(k_N+k_M+x_N+x_M)\end{math}, and \begin{math}x_{NM}\end{math} be the corresponding fractional part. Then, the exact product can be simplified without considering carry propagation:
\begin{subequations}\label{eq:12}
\begin{align}
NM&=2^{k_{NM}}(1+x_{NM})+2^{k_N+k_M}(x_Nx_M),\ x_N+x_M<1 \label{eq:12A}\\
NM&=2^{k_{NM}}(1+x_{NM})+2^{k_N+k_M}(x_N^\prime x_M^\prime),\ x_N+x_M\geq1 \label{eq:12B}
\end{align}
\end{subequations}

By recursively calculate the product terms \begin{math}x_Nx_M\end{math} and \begin{math}x_N^\prime x_M^\prime\end{math} using (\ref{eq:12A}) and (\ref{eq:12B}), arbitrary precision of the product can be achieved. However, such nesting leads to excessive chip area consumption and computational latency. In practice, the correction terms are often approximated using (\ref{eq:9A}) or (\ref{eq:9B}), which reduces the maximum error to about 2.8\% while preserving high calculation speed. Since the combined error behavior of Mitchell and Goldschmidt algorithms is difficult to analyses theoretically, error analysis for the proposed divider is provided in Section IV through empirical evaluation.

Additionally, although Mitchell algorithm can directly approximate the division by replacing addition with subtraction, the accuracy is limited to 87.5\%, and no correction term can effectively improve the precision. Therefore, this approach is generally not adopted.

\section{FPGA-Based Implementation}
\subsection{Overall Hardware Architecture}
\begin{figure}[h]
\centering
\includegraphics[width=3in]{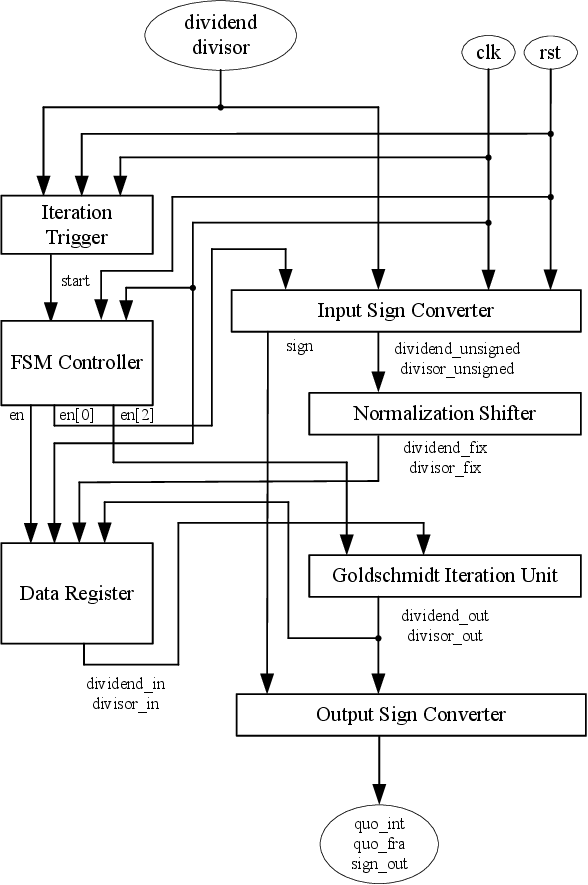}
\caption{The overall hardware architecture of the proposed divider.}
\label{fig_1}
\end{figure}

The overall hardware architecture of the proposed divider is shown in Fig. \ref{fig_1}. It consists of an input sign converter, an iteration trigger, a finite state machine (FSM) controller, a data register, a normalization shifter, a Goldschmidt iteration unit, which incorporates Mitchell multiplier units and adders, as well as an output sign converter.

In Fig. \ref{fig_1}, the dividend, divisor, clock signal (clk), and reset signal (rst) are the input signals. The signals dividend\_unsigned and divisor\_unsigned are the absolute value of dividend and divisor. The signal sign is the sign of the quotient. The signal start is the trigger signal initiating one division iteration. The signals en[3-0] are the enable signals. The signals dividend\_fix, divisor\_fix, dividend\_out and divisor\_out are the intermediate results of the iterations.

For convenience of description, the algorithm flow discussed in the following sections assumes that the enable signal en is active and rst is inactive. When the divisor is zero, the computation is considered invalid, and the output data is undefined.

\subsection{Input Sign Converter}
\begin{figure}[h]
\centering
\includegraphics[width=2.5in]{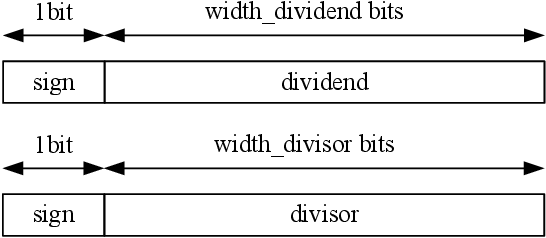}
\caption{The data formats of input signals.}
\label{fig_2}
\end{figure}

The input signals dividend and divisor are signed integers, where the bit-widths are parameterized as $width\_dividend$ and $width\_divisor$, respectively. Their data formats are shown in Fig. \ref{fig_2}. 

\begin{figure}[h]
\centering
\includegraphics[width=2.5in]{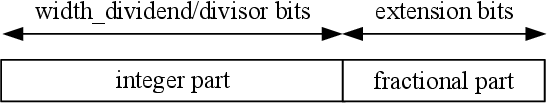}
\caption{The data formats of internal signals.}
\label{fig_3}
\end{figure}

As shown in Fig. \ref{fig_3}, the format of the input sign converter’s output is the unsigned fixed-point, in which the bit-width of the fractional part is defined by parameter $extension$, while the bit-width of the integer part is still $width\_dividend$ or $width\_divisor$. The same format is also employed for the intermediate results dividend\_fix, divisor\_fix, dividend\_out and divisor\_out. 

The input sign converter is implemented using combinational logic, and its algorithm consists of the following steps:
\begin{itemize}
\item The input signals dividend and divisor are converted to their absolute value according to their sign bits, and output as dividend\_unsigned or divisor\_unsigned;
\item The sign bits of dividend and divisor are XORed, and the result is output as the signal sign.
\end{itemize}

\subsection{Iteration Trigger}
The iteration trigger is implemented using sequential logic. Its algorithm contains following steps:
\begin{itemize}
\item The input signals dividend and divisor are sent into two cascaded D flip-flops driven by the rising edge of clk, producing outputs dividend\_1 and divisor\_1, respectively;
\item If dividend = dividend\_1 and divisor = divisor\_1, the output signal start is set to 0; otherwise, start is set to 1.
\end{itemize}

\subsection{FSM Controller}
\begin{figure}[h]
\centering
\includegraphics[width=3.3in]{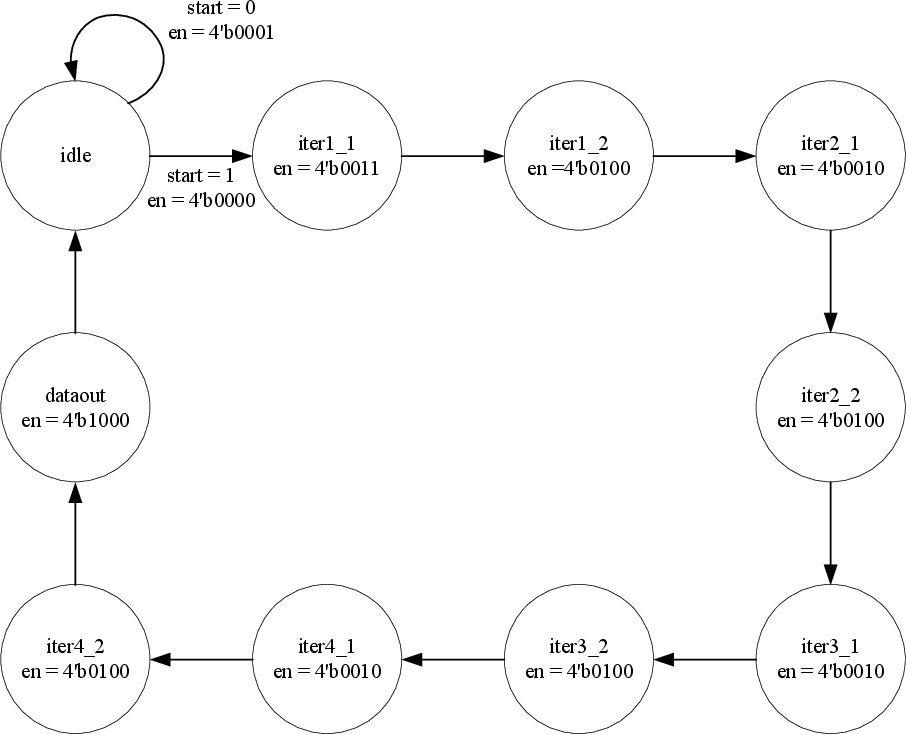}
\caption{The state transition diagram of the FSM controller.}
\label{fig_4}
\end{figure}

The FSM controller is implemented using sequential logic, and its state transition diagram is shown in Fig. \ref{fig_4}. The number of iterations is set to four, which represents a trade-off between computational accuracy and latency. The FSM controller generates the enable signals en[3-0] (active-high) and controls the data flow of the data register, thereby governing the progress of the computation. The function of each bit of en is shown as follows:
\begin{itemize}
\item When en[0]=1, the input sign converter is enabled and stays ready while waiting for input data, then performs the sign conversion in the first clock cycle after data arrival;
\item When en[1]=1, the adder calculates and outputs the iteration coefficient;
\item When en[2]=1, the multiplier unit is enabled and performs the multiplication operation;
\item When en[3]=1, the output sign converter outputs the final result.
\end{itemize}

\subsection{Normalization Shifter}
The normalization shifter is implemented using combinational logic, and its algorithm can be summarized as the following steps:
\begin{itemize}
    \item A flag signal (once) is initialized to 1, and the intermediate value $shift\_length\_R$ is initialized to 0;
    \item When the input signal divisor\_unsigned changes, a loop is initiated. While once = 1, the bits of divisor\_unsigned are checked sequentially from the most significant bit (MSB) to the least significant bit (LSB). If the current bit is 0, the value of $shift\_length\_R$ is added by 1. If the current bit is 1, the flag signal (once) is set to 0, then no further operations are performed in the remaining loop;
    \item Finally, divisor\_unsigned is left-shifted by $(extension - width\_divisor + shift\_length\_R - 1)$ bits to obtain divisor\_fix, which falls within the interval [0.5,1). Then dividend\_unsigned is left-shifted by the same amount to align with divisor\_fix, yielding dividend\_fix.
\end{itemize}

\subsection{Goldschmidt Iteration Unit}
\begin{figure}[h]
\centering
\includegraphics[width=3.1in]{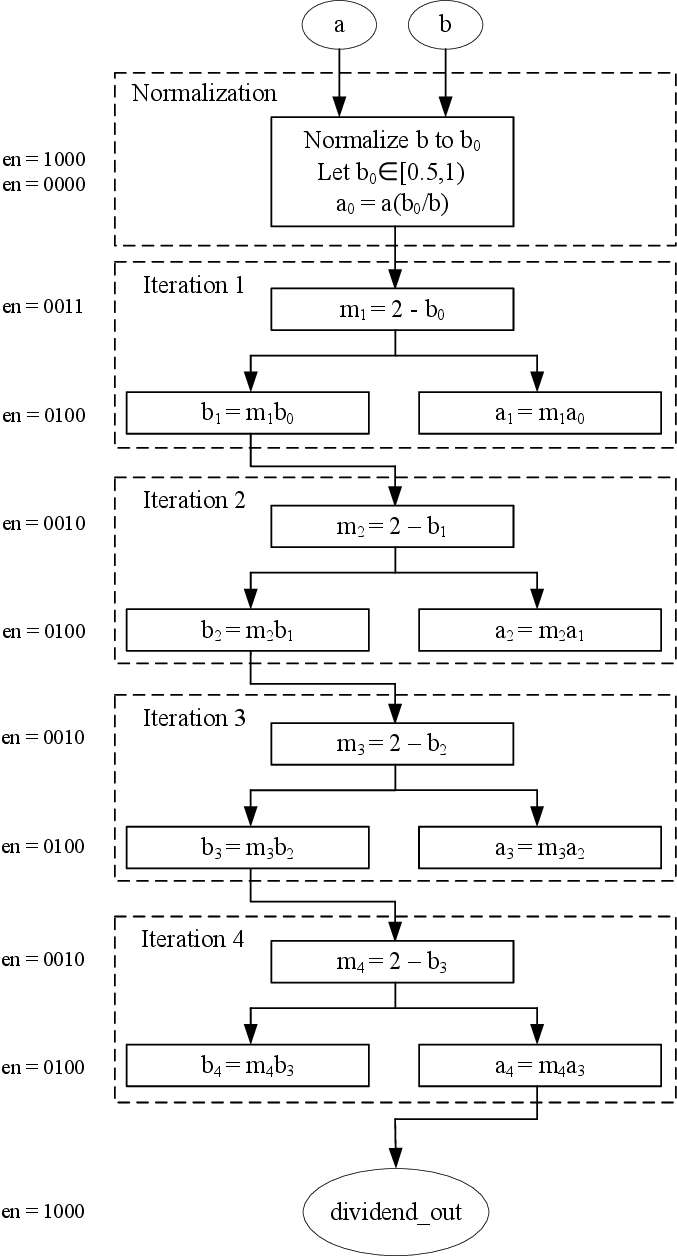}
\caption{The iteration process of Goldschmidt algorithm.}
\label{fig_5}
\end{figure}

The iteration process of the Goldschmidt algorithm is illustrated in Fig. \ref{fig_5}. For simplicity, $a$ and $b$ represent dividend\_unsigned and divisor\_unsigned, which are the output of the input sign converter. $a_0$ and $b_0$ represent dividend\_fix and divisor\_fix, which are the output of the normalization shifter. $m$ represents the iteration coefficient.

Since the proposed divider employs a data register to store intermediate results, the iteration unit can be reused across iterations. As a result, the hardware implementation requires only one adder and two multiplier units.

Goldschmidt algorithm can be summarized as following steps:
\begin{itemize}
    \item After normalization, the initial divisor \begin{math}b_0\end{math} is obtained within the interval [0.5,1). The initial dividend \begin{math}a_0\end{math} is obtained by aligning with \begin{math}b_0\end{math};
    \item In each iteration, the iteration coefficient \begin{math}m_k\end{math} is first computed, followed by parallel updates of the numerator and denominator;
    \item After four iterations, the value of \begin{math}a_4\end{math} is output as the approximate quotient.
\end{itemize}

\subsection{Data Register}
The data register is implemented using sequential logic, triggered by the rising edge of clk. Its function is to store the intermediate results of the Goldschmidt iterations, including dividend\_fix, divisor\_fix, dividend\_out, and divisor\_out. Depending on the value of the enable signal en, the register selectively loads different intermediate data into dividend\_in and divisor\_in at different computation stages, which are then sent to the iteration unit for the subsequent iteration step.

\begin{figure}[h]
\centering
\includegraphics[width=3.3in]{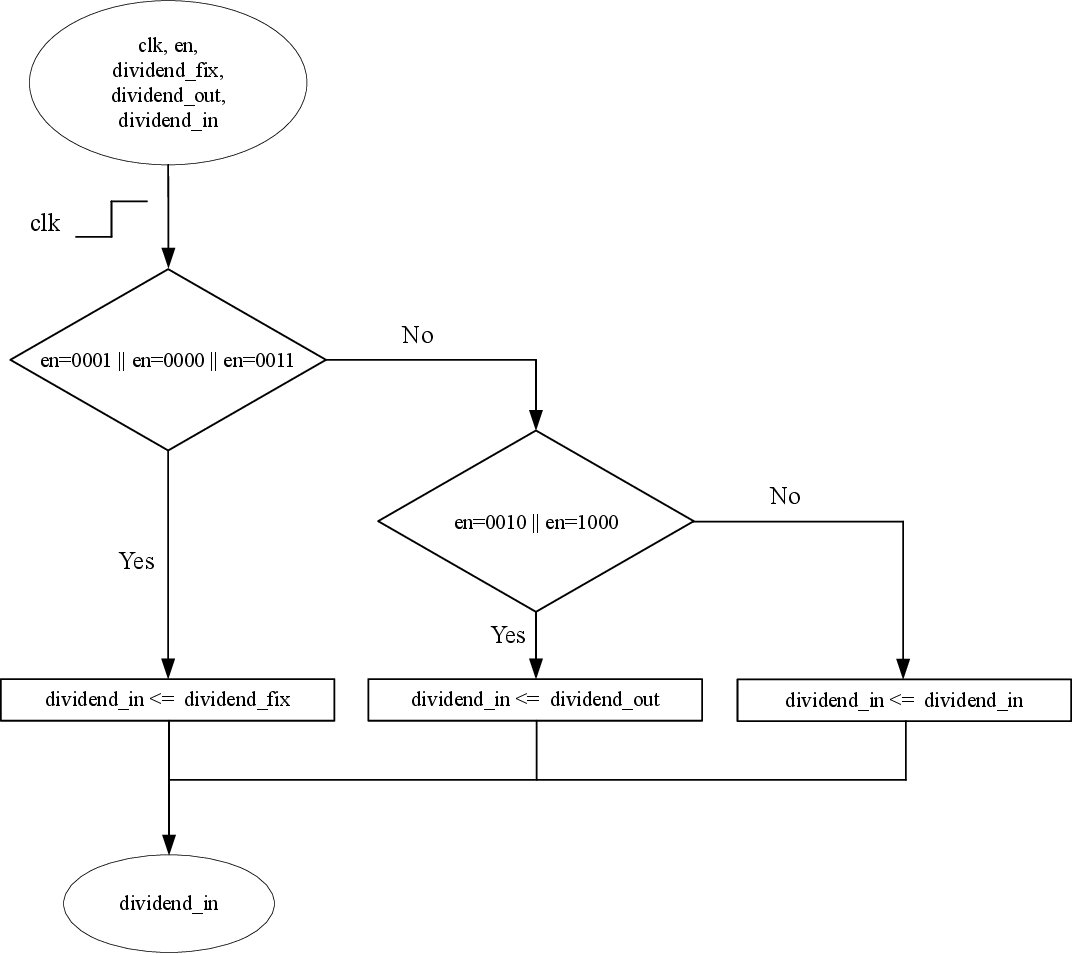}
\caption{The algorithm of the data register.}
\label{fig_6}
\end{figure}

The algorithm of the data register is illustrated in Fig. \ref{fig_6}. Since the logic for the divisor is identical to that for the dividend, only the algorithm for the dividend is presented.

\subsection{Mitchell Multiplier Unit}
\begin{figure*}[htbp]
\centering
\includegraphics[width=4.6in]{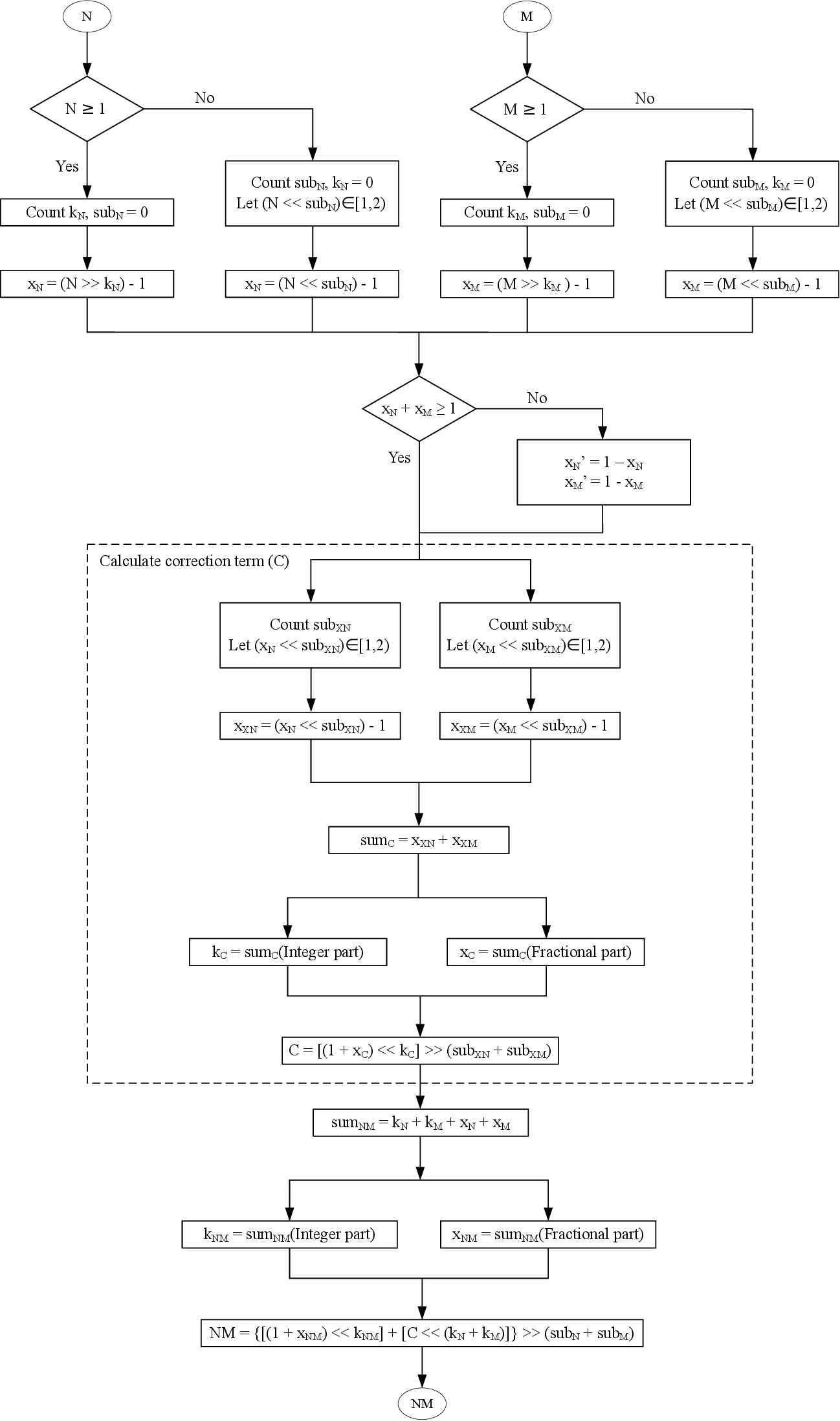}
\caption{The algorithm of the Mitchell multiplier unit with correction.}
\label{fig_7}
\end{figure*}

Based on the principles discussed in Section II, the Mitchell multiplier unit with correction, as illustrated in Fig. \ref{fig_7}, is implemented using combinational logic and controlled by the signal en. Each multiplier unit consists of a primary multiplier, a correction multiplier, and two auxiliary shifters. Let the two operands be denoted as $N$ and $M$.

It is important to note that the equations in Section II are valid only when both $N$ and $M$ are greater than 1. If either operand is less than 1, it should be left-shifted by $sub_N$ or $sub_M$ bits to ensure the values fall within the interval [1,2). After computation, the result is right-shifted by $(sub_N + sub_M)$ bits to restore the correct magnitude.

The operations Count k and Count sub in Fig. \ref{fig_7} consist of the following steps:
\begin{itemize}
    \item The operands $N$,$M$,\begin{math}x_N\end{math},\begin{math}x_M\end{math} are sent to the auxiliary shifters, which generate the corresponding $shift\_length$ values.
    \item The value of $k$ and $sub$ are then computed according to:
\end{itemize}
\begin{equation}
\label{deqn_ex13}
k=shift\_length-extension-1,
\end{equation}
\begin{equation}
\label{deqn_ex14}
sub=extension-shift\_length+1.
\end{equation}

During the computation of the correction term C, it is observed that \begin{math}x_N\end{math},\begin{math}x_M\end{math},\begin{math}x_N^\prime\end{math},\begin{math}x_M^\prime\end{math}$<1$. Therefore, the branch of calculating $k$ can be omitted.

\subsection{Auxiliary Shifter}
The auxiliary shifter is implemented using combinational logic. Let the two operands of the multiplication be denoted as num1 and num2. Its algorithm can be summarized as following steps:
\begin{itemize}
    \item The algorithm of the normalization shifter is performed on the operands num1 and num2, obtaining the corresponding intermediate values $shift\_length\_num\_R\_1$ and $shift\_length\_num\_R\_2$;
    \item The value of $shift\_length\_num$ for each operand is calculated according to the following expression:
\end{itemize}
\begin{align}
shift\_length\_num=&width\_num+extension–\notag \\
&shift\_length\_num\_R+1.
\label{deqn_ex15}
\end{align}

Then $shift\_length\_num$ is sent to the multiplier for the subsequent calculation of $k$ and $sub$.

\subsection{Output Sign Converter}
\begin{figure}[h]
\centering
\includegraphics[width=2.5in]{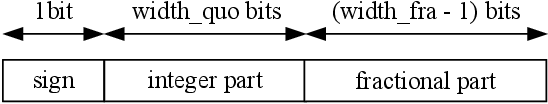}
\caption{The data format of the final result.}
\label{fig_8}
\end{figure}

The final quotient of the divider is represented as a signed fixed-point number, as shown in Fig. \ref{fig_8}. The bit-width of the integer part is defined by the parameter $width\_quo$, while that of the fractional part is defined by the parameter $(width\_fra-1)$.

The output sign converter is implemented using combinational logic. According to the signal sign, it converts the final iteration result dividend\_out into the corresponding signed number, which is the final quotient. The final quotient is then partitioned into the integer part quo\_int and the fractional part quo\_fra, and they are output separately. Additionally, the sign bit of the quotient is also output as the signal sign\_out.

\section{On-Board Operation Results and Performance Analysis}
The proposed divider is described in Verilog HDL and implemented on a Xilinx XC7Z020-2CLG400I FPGA. For on-board testing, the bit-width of input is configured to 32 bits (single-precision), and the quotient is output in signed fixed-point format with 32-bit integer and 32-bit fractional parts. Calculation examples and the corresponding error analysis are shown in Table \ref{tab:table1}.
\begin{table}[h]
\caption{Calculation Examples and Error Analysis\label{tab:table1}}
\centering
\begin{tabular}{|c|c|c|c|c|}
\hline
Dividend & Divisor & \makecell{Accurate\\ Quotient} & \makecell{Computed\\ Quotient} & \makecell{Relative\\ Error(\%)}\\
\hline
$-17$ & $35$ & $-0.4857$ & $-0.4868$ & $0.22$\\
\hline
$53$ & $11$ & $4.8181$ & $4.8253$ & $0.14$\\
\hline
$345$ & $4252$ & $0.0811$ & $0.0812$ & $0.10$\\
\hline
$2741$ & $67$ & $40.9104$ & $40.9342$ & $0.05$\\
\hline
$34242$ & $5567$ & $6.1508$ & $6.1759$ & $0.40$\\
\hline
$89230293$ & $432424$ & $206.3490$ & $206.7367$ & $0.18$\\
\hline
$2^{31}-1$ & $947483647$ & $2.2665$ & $2.2685$ & $0.08$\\
\hline
$2^{31}-1$ & $-47483647$ & $-45.2257$ & $-45.4989$ & $0.60$\\
\hline
\end{tabular}
\end{table}

More than 100 tests were performed, confirming that the relative error of the computed quotient is below 1\%.

\begin{figure*}[htbp]
\centering
\includegraphics[width=7in]{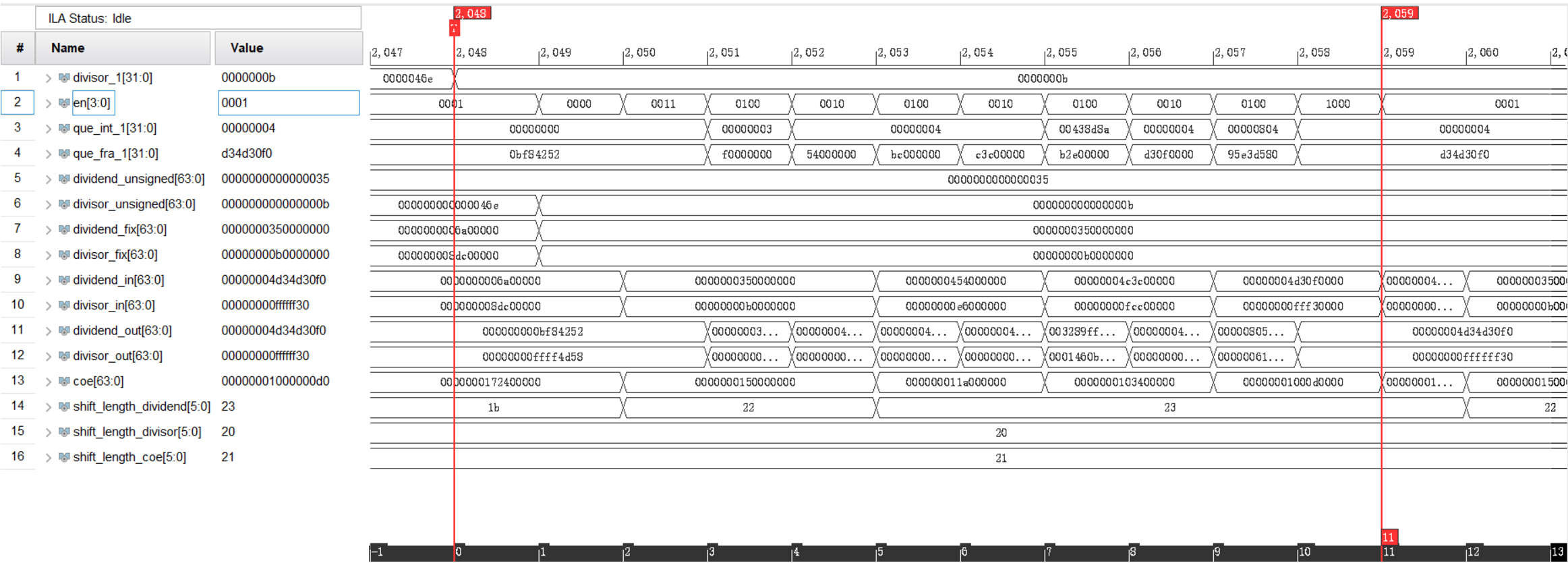}
\caption{The waveform of the on-board calculation.}
\label{fig_9}
\end{figure*}

The timing test was conducted using the FPGA’s Virtual I/O (VIO) core. For the input $53/11$, the waveform in Fig. \ref{fig_9} shows a total latency of 11 clock cycles from trigger to completion.

The Xilinx implemented power report of the proposed divider is shown in Table \ref{tab:table2}.
\begin{table}[h]
\caption{An Example of a Table\label{tab:table2}}
\centering
\begin{tabular}{|c||c|}
\hline
Specifications & Values\\
\hline
Total On-Chip Power & $0.152W$\\
\hline
Junction Temperature & $26.7^\circ C$\\
\hline
Thermal Margin & $58.3^\circ C(4.9W)$\\
\hline
Effective $\theta JA$ & $11.5^\circ C/W$\\
\hline
\end{tabular}
\end{table}

A comparison of latency and operating speed with existing dividers is presented in Table \ref{tab:table3}.
\begin{table}[htb]
\caption{An Example of a Table\label{tab:table3}}
\centering
\begin{tabular}{|c||c|c|c|c|}
\hline
Algorithm & \makecell{Setup\\ WNS(ns)} & \makecell{fmax\\ (fmax=\\ 1/(T-WNS))\\ (MHz)} & \makecell{Number of\\ clock cycles\\ required} & \makecell{Latency\\ time(ns)}\\
\hline
O-FixDiv\cite {ref5} & \makecell{$3.997$\\ $(@70MHz)$} & $97$ & $Q$ & $10.289*Q$\\
\hline
RS-FixDiv\cite {ref6} & \makecell{$4.619$\\ $(@100MHz)$} & $186$ & $Q$ & $5.381*Q$\\
\hline
FloatDiv\cite{ref7} & $-$ & $-$ & $-$ & $130.8$\\
\hline
\makecell{Goldschmidt\\ FloatDiv\\ (Vedic) \cite{ref3}} & $-$ & $-$ & $-$ & $75$\\
\hline
This work & \makecell{$11.068$\\ $(@50MHz)$} & $111$ & $11$ & $99.1$\\
\hline
\end{tabular}
\end{table}

Resource utilization comparison with another Goldschmidt-based divider \cite{ref3}, are shown in Table \ref{tab:table4}.
\begin{table}[h]
\caption{An Example of a Table\label{tab:table4}}
\centering
\begin{tabular}{|c||c|c|c|c|}
\hline
\ & Slice Registers & Slice LUTs & Slices & BUFG\\
\hline
Ref. [3] & $3025$ & $9281$ & $2961$ & $-$\\
\hline
This work & $1613$ & $8823$ & $2610$ & $3$\\
\hline
\end{tabular}
\end{table}

The major drawback of existing fixed-point dividers \cite{ref5}, \cite{ref6} is that the computation latency depends on the quotient Q. As Q increases, the required computation latency becomes longer, which introduces uncertainty and complicates the design of other modules that rely on the divider. In contrast, the minimum computation latency of the proposed divider is fixed at 99.1 ns, thereby avoiding the long-latency issue of dividers in \cite{ref5} and \cite{ref6} when Q is large. Furthermore, the proposed design achieves a speed advantage of 31.7 ns compared with the existing single-precision floating-point divider \cite{ref7}.

Within the same Goldschmidt algorithm framework, a comparison between the proposed divider using Mitchell multiplier units and the divider based on Vedic multiplier units \cite{ref3} shows that the proposed divider reduces Slice Registers by 46.68\%, Slice LUTs by 4.93\%, and Slices by 11.85\%, while the computation latency increases by only 24.1 ns and the relative error increases by less than 1\%. Overall, the proposed divider achieves a more favorable trade-off between latency and hardware resource utilization.

\section{Conclusion}
In this paper, a variable bit-width fixed-point fast divider using Goldschmidt division algorithm and Mitchell multiplication algorithm has been proposed, implemented, and tested on the Xilinx XC7Z020-2CLG400I FPGA platform. The proposed divider is designed with parameterized input and output bit-widths, which enables flexible application in different computational systems. On-board test results demonstrate that the proposed divider achieves more than 99\% accuracy, with a minimum computation latency of 99.1 ns, which is 31.7 ns faster than the existing single-precision floating-point divider \cite{ref7}. Compared with existing fixed-point dividers \cite{ref5}, \cite{ref6}, the proposed architecture achieves both shorter and fixed computation latency, eliminating the performance uncertainty associated with the value of the quotient. Furthermore, the proposed divider saves 46.68\% Slice Registers, 4.93\% Slice LUTs, and 11.85\% Slices compared with the Goldschmidt divider using Vedic multiplier \cite{ref3}, at the cost of additional 24.1 ns latency and less than 1\% relative error. The proposed divider is particularly well-suited for applications requiring high-speed computation with constrained hardware resources, such as image processing, industrial control, and Internet of Things (IoT) devices.

\end{document}